\documentclass[letterpaper, 10 pt, conference]{ieeeconf}
\usepackage{amsmath,amsfonts,amssymb}
\usepackage{graphicx}
\usepackage{BernsteinStyle}

\IEEEoverridecommandlockouts

\title{\LARGE \bf
Fast RLS Identification Leveraging the Linearized System Sparsity: Predictive Cost Adaptive Control for Quadrotors$^*$
\thanks{$^*$Preprint submitted to the American Control Conference (ACC) 2026.}
}

\author{Tam W. Nguyen$^{1}$
\thanks{$^{1}$Tam W. Nguyen is with the Department of Electrical Engineering, Kyoto University, Kyoto 615-8510, Japan {\tt\small nguyen.tamwilly.3e@kyoto-u.ac.jp}}%
}

\begin{document}

\maketitle
\thispagestyle{empty}
\pagestyle{empty}

\begin{abstract}

This paper presents a centralized predictive cost adaptive control (PCAC) strategy for the position and attitude control of quadrotors.
PCAC is an optimal, prediction-based control method that uses recursive least squares (RLS) to identify model parameters online, enabling adaptability in dynamic environments.
Addressing challenges with black-box approaches in systems with complex couplings and fast dynamics, this study leverages the unique sparsity of quadrotor models linearized around hover points.
By identifying only essential parameters related to nonlinear couplings and dynamics, this approach reduces the number of parameters to estimate, accelerates identification, and enhances stability during transients. 
Furthermore, the proposed control scheme removes the need for an attitude setpoint, typically required in conventional cascaded control designs.

\end{abstract}

\section{Introduction}

Unmanned Aerial Vehicles (UAVs), particularly quadrotors, have become essential in various applications due to advancements in sensor technology and increasingly affordable hardware \cite{rs11121443}. These developments have spurred extensive research in autonomous quadrotor control \cite{8827409, idrissi2022review}, driven by their versatility in civil and military fields \cite{6196930, aerospace8120363}.

Quadrotor control architectures commonly use a cascaded design, with an outer-loop position controller generating attitude setpoints, and a faster inner-loop directly controlling the attitude \cite{szafranski2011different, PAREDES2021372}. These inner-outer loops rely on proportional-integral-derivative (PID) controllers, with control gains tuned through simulations and refined experimentally. However, in tilting multirotor systems, where attitude generation requires greater complexity \cite{9144371,NGUYEN2021109586}, a centralized approach, where a single unit manages both inner and outer loops, offers potential advantages.
Fixed control gains may suit specific models in controlled environments but often need adaptation when external factors, such as variable payloads or model uncertainties, affect quadrotor dynamics \cite{WANG2023108088}.

This paper explores a centralized approach with predictive cost adaptive control (PCAC) \cite{9612636, 9482724} for the position and attitude control of quadrotors.
PCAC is an optimal, prediction-based control method that uses recursive least squares (RLS) to continuously identify model parameters online, enabling responsiveness to changing conditions.
Recent extensions of PCAC in flight control have been investigated in \cite{doi:10.2514/6.2025-0033, doi:10.2514/6.2025-2081, doi:10.2514/1.G008859}.
While PCAC traditionally employs a general input-output model, this study leverages the inherent sparsity in quadrotor state-space models linearized around hover points.
By identifying only key parameters associated with the system nonlinear couplings and rapid dynamics, this approach reduces the number of parameters to identify, accelerates the identification process, and enhances stability during transients.

The paper is structured as follows:
Section \ref{sec:modeling} describes the quadrotor dynamics, which are linearized around hover points in Section \ref{sec:linearization}, highlighting the unique sparsity of the matrices.
Section \ref{sec:sparsity} leverages this sparsity to identify only essential parameters.
Section \ref{sec:pcac} introduces PCAC, tailored specifically to quadrotors, and Section \ref{sec:simulations} demonstrates PCAC's robustness to varying initial conditions and its resilience to abrupt changes in the quadrotor mass.
Finally, Section \ref{sec:conclusion} summarizes the results and outlines directions for future works.

\section{Modeling}\label{sec:modeling}

\begin{figure}
    \centering
    \includegraphics[width=0.75\columnwidth]{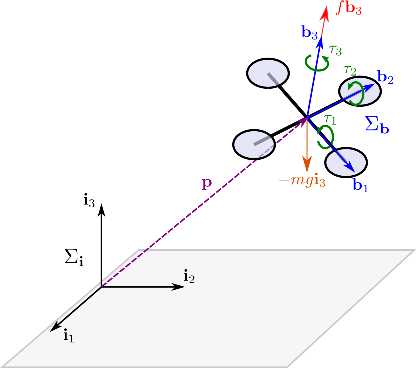}
    \caption{Quadrotor system. The inertial frame is denoted by $\Sigma_{\vec{i}} = \{\vec{i}_1,\vec{i}_2,\vec{i}_3\},$ and the body-fixed frame by $\Sigma_{\vec{b}} = \{\vec{b}_1,\vec{b}_2,\vec{b}_3\}.$
    The quadrotor, whose position is denoted by $\vec{p},$ is subject to the gravity force $-mg\vec{i}_3,$ and actuated by the thrust $f\vec{b}_3$ and torque $\vec{\tau}=[\tau_1 \; \tau_2 \; \tau_3]^\rmT.$
    }
    \label{fig:quadrotor}
\end{figure}

Consider the quadrotor shown in Figure~\ref{fig:quadrotor}, where the inertial frame is denoted by $\Sigma_{\vec{i}}=\{\vec{i}_1,\vec{i}_2,\vec{i}_3\}$ and the body-fixed frame by $\Sigma_{\vec{b}}=\{\vec{b}_1,\vec{b}_2,\vec{b}_3\}.$
The vectors $\vec{i}_j\in\BR^3$ and $\vec{b}_j\in\BR^3$ are unit basis vectors of $\Sigma_{\vec{i}}$ and $\Sigma_{\vec{b}},$ respectively, where $j = 1,$ $j=2,$ and $j=3$ correspond to the $x$-, $y$-, and $z$-basis vectors, respectively.

Let $\vec{p}\in\BR^3$ be the position of the quadrotor in $\Sigma_{\vec{i}},$ $R\in SO(3)$ the rotation matrix from $\Sigma_{\vec{b}}$ to $\Sigma_{\vec{i}},$ and $\vec{\omega}\in\BR^3$ the angular velocity in $\Sigma_{\vec{b}}.$
Furthermore, let $f\in\BR_{\ge0}$ be the total thrust along $\vec{b}_3,$ and $\vec{\tau}\in\BR^3$ the total torque in $\Sigma_{\vec{b}}.$
The equations of motion of the quadrotor are
\begin{align}
    m\ddot{\vec{p}} &= -mg\vec{e}_3 + fR^\rmT \vec{e}_3, \label{eq:eom1} \\
    \dot{R} &= R\vec{\omega}^\times, \label{eq:eom2} \\
    J \dot{\vec{\omega}} &= -\vec{\omega}\times J\vec{\omega} + \vec{\tau}, \label{eq:eom3}
\end{align}
where $m\in\BR_{>0}$ is the total quadrotor mass, $J\in\RR{3}_{>0}$ the inertia matrix with respect to $\Sigma_{\vec{b}},$ $g\in\BR_{>0}$ the gravitational acceleration, and $\vec{e}_3\triangleq [0 \; 0 \; 1]^\rmT.$
Given $\vec{\omega} = [\omega_1 \; \omega_2 \; \omega_3]^\rmT,$ the skew operator $(\cdot)^\times\colon\BR^3\to so(3)$ is defined by
\begin{align}
    \left[
    \begin{matrix}
        \omega_1 \\ \omega_2 \\ \omega_3
    \end{matrix}
    \right]^\times \triangleq
    \left[
    \begin{matrix}
        0 & -\omega_3 & \omega_2 \\
        \omega_3 & 0 & -\omega_1 \\
        -\omega_2 & \omega_1 & 0
    \end{matrix}
    \right].
\end{align}


Next, $R$ is parameterized by the Euler angles following the $Z\to X \to Y$ sequence.
Specifically, rotate first by the yaw angle $\psi$ about $\vec{b}_3,$ then the roll angle $\phi$ about $\vec{b}_1,$ and finally the pitch angle $\theta$ about $\vec{b}_2.$
This rotation aligns $\Sigma_{\vec{b}}$ with $\Sigma_{\vec{i}}.$
Let $R_x,$ $R_y,$ and $R_z$ be the rotation matrices about the $x$-, $y$-, and $z$-axes, respectively.
The rotation matrix $R$ can be computed by
\begin{align}
    R & = R_y R_x R_z \nonumber \\
    & =
    \left[
    \begin{smallmatrix}
    \cos{\theta} & 0 & \sin{\theta} \\
    0 & 1 & 0 \\
    -\sin{\theta} & 0 & \cos{\theta}
    \end{smallmatrix}
    \right]
    \left[
    \begin{smallmatrix}
        1 & 0 & 0 \\
        0 & \cos{\phi} & -\sin{\phi} \\
        0 & \sin{\phi} & \cos{\phi}
    \end{smallmatrix}
    \right]
    \left[
    \begin{smallmatrix}
        \cos{\psi} & -\sin{\psi} & 0 \\
        \sin{\psi} & \cos{\psi} & 0 \\
        0 & 0 & 1
    \end{smallmatrix}
    \right]
    \nonumber \\
    & = \left[
    \begin{matrix}
    c_\psi c_\theta + s_\phi s_\psi s_\theta & c_\psi s_\phi s_\theta - c_\theta s_\psi & c_\phi s_\theta \\
    c_\phi s_\psi & c_\phi c_\psi & -s_\phi \\
    c_\theta s_\phi s_\psi - c_\psi s_\theta & s_\psi s_\theta + c_\psi c_\theta s_\phi & c_\phi c_\theta
    \end{matrix}
    \right],
\end{align}
where $c_\theta$ and $s_\theta$ denote $\cos{\theta}$ and $\sin{\theta},$ respectively, with a similar convention for $\phi$ and $\psi.$

Let $\vec{e}_1 \triangleq [1 \; 0 \; 0]^\rmT$ and $\vec{e}_2 \triangleq [0 \; 1 \; 0]^\rmT.$
The angular velocity $\vec{\omega}$ can be expressed by
\begin{align}
    \vec{\omega} & = \dot{\psi} \vec{e}_3 + \dot{\phi} R_z \vec{e}_1 + \dot{\theta} R_xR_z \vec{e}_2 \nonumber \\
    & = \underbrace{
    \left[
    \begin{matrix}
    0 & \cos{\psi} & -\sin{\psi} \\
    0 & \sin{\psi} & \cos{\phi}\cos{\psi} \\
    1 & 0 & \sin{\phi}\cos{\psi}
    \end{matrix}
    \right]
    }_{J_{\vec{\omega}}}
    \dot{\vec{\xi}},
    \label{eq:Omega_euler}
\end{align}
where $\vec{\xi} \triangleq [\psi \; \phi \; \theta]^\rmT$ is the Euler-angle vector.
By inverting the Jacobian $J_{\vec{\omega}}$ in \eqref{eq:Omega_euler}, the Euler kinematics are obtained by
\begin{align}
    \dot{\vec{\xi}} = J_{\vec{\omega}}^{-1} \vec{\omega} = \frac{1}{c^2_{\psi} c_\phi + s^2_\psi} 
    \left[
    \begin{matrix}
    s_\phi c_\psi s_\psi & -s_\phi c^2_\psi & c^2_\psi c_\phi + s^2_\psi \\
    c_\phi c_\psi & s_\psi & 0 \\
    - s_\psi & c_\psi & 0
    \end{matrix}
    \right]
    \vec{\omega}. \label{eq:euler_kinematics}
\end{align}

\section{Linearization Near Hover Points}\label{sec:linearization}

The hover points are the equilibria calculated by zeroing the time derivatives of \eqref{eq:eom1}-\eqref{eq:eom3}.
The hover points are any $\vec{p}\in\BR^3,$ where $R = I_3,$ $\vec{\omega} = [0 \; 0 \; 0]^\rmT,$ $\vec{\tau} = [0 \; 0 \; 0]^\rmT,$ and $f = mg,$ where $I_n\in\RR{3}$ is the $n$-by-$n$ identity matrix.

Define the state $x \triangleq [\vec{p}^\rmT \; \vec{\xi}^\rmT \; \dot{\vec{p}}^\rmT \; \vec{\omega}^\rmT]^\rmT\in\BR^{12}$ and input $u\triangleq[f \; \vec{\tau}^\rmT]^\rmT\in\BR^{4}.$
The equations of motion \eqref{eq:eom1}-\eqref{eq:eom3} using the Euler kinematics \eqref{eq:euler_kinematics} are rewritten as
\begin{align}
    \dot{x} = \left[
    \begin{matrix}
        \dot{\vec{p}} \\
        J_{\vec{\omega}}^{-1} \vec{\omega} \\
        -g\vec{e}_3 + \frac{f}{m}R^\rmT\vec{e}_3 \\
        J^{-1}(-\vec{\omega} \times J\vec{\omega} + \vec{\tau})
    \end{matrix}
    \right] = \SF(x,u). \label{eq:nonlinear_system}
\end{align}

Next, define the equilibrium state $\bar{x} \triangleq [\vec{p}^\rmT \; 0_{1\times9}]^\rmT\in\BR^{12}$ and equilibrium input $\bar{u} \triangleq [mg \; 0_{1\times3}]^\rmT\in\BR^{4},$ where $0_{m\times n}$ denotes the $m$-by-$n$ matrix of zeros. 
Let $\tilde{x} = x - \bar{x}$ and $\tilde{u} = u - \bar{u}.$
The system linearized around $\bar{x}$ and $\bar{u}$ is
\begin{align}
    \dot{\tilde{x}} = & 
    \underbrace{
    \left[
    \left.
    \begin{matrix}
        \frac{\partial \SF_{1}}{\partial x_{1}} & \ldots & \frac{\partial \SF_{1}}{\partial x_{12}} \\
        \vdots & \ddots & \vdots \\
        \frac{\partial \SF_{12}}{\partial x_1} & \ldots & \frac{\partial \SF_{12}}{\partial x_{12}}
    \end{matrix}
    \right]\right|_{(\bar{x}, \bar{u})} 
    }_{A}
    \tilde{x} + \nonumber \\ 
    &\underbrace{
    \left[
    \left.
    \begin{matrix}
        \frac{\partial \SF_{1}}{\partial u_{1}} & \frac{\partial \SF_{1}}{\partial u_{2}} & \frac{\partial \SF_{1}}{\partial u_{3}} & \frac{\partial \SF_{1}}{\partial u_{4}} \\
        \vdots & \vdots & \vdots & \vdots \\
        \frac{\partial \SF_{12}}{\partial u_{1}} & \frac{\partial \SF_{12}}{\partial u_{2}} & \frac{\partial \SF_{12}}{\partial u_{3}} & \frac{\partial \SF_{12}}{\partial u_{4}}
    \end{matrix}
    \right]\right|_{(\bar{x}, \bar{u})} 
    }_{B}
    \tilde{u},
\end{align}
where $\SF_i\colon\BR^{12}\times\BR^{4}\to\BR,$ $x_i\in\BR,$ and $u_i\in\BR$ are the $i$-th component of $\SF,$ $x,$ and $u,$ respectively. The matrices $A\in\RR{12}$ and $B\in\BR^{12\times4}$ are
\begin{align}
    A & =
    \left[
    \begin{smallmatrix}
         0_{3\times4} & 0_{3\times1} & 0_{3\times1} & I_3 & 0_{3\times1} & 0_{3\times1} & 0_{3\times1} \\
         0_{1\times4} & 0 & 0 & 0_{1\times3} & 0 & 0 & 1 \\
         0_{1\times4} & 0 & 0 & 0_{1\times3} & 1 & 0 & 0 \\
         0_{1\times4} & 0 & 0 & 0_{1\times3} & 0 & 1 & 0 \\
         0_{1\times4} & 0 & -g & 0_{1\times3} & 0 & 0 & 0 \\
         0_{1\times4} & g & 0 & 0_{1\times3} & 0 & 0 & 0 \\
         0_{4\times4} & 0_{4\times1} & 0_{4\times1} & 0_{4\times3} & 0_{4\times1} & 0_{4\times1} & 0_{4\times1}
    \end{smallmatrix}
    \right], \nonumber \\
    B & = \left[
    \begin{matrix}
         0_{8\times1} & 0_{8\times3} \\
         m^{-1} & 0_{1\times3} \\
         0_{3\times1} & J^{-1} \\
    \end{matrix}
    \right]. \label{eq:linearized_system}
\end{align}
Note that, for all nonzero $g,$ $m,$ and $J,$ the controllability matrix has rank 12, rendering the linearized system controllable.
Furthermore, in the following, we assume $x$ measurable, which is reasonable and well-justified in most UAV systems.

\section{Leveraging the Linearized System Sparsity in RLS}\label{sec:sparsity}

Note that $A$ and $B$ in \eqref{eq:linearized_system} are sparse, with 94.4\% and 91.7\% sparsity, respectively.
Furthermore, note that all elements of $A$ are reasonably and accurately known in advance.
However, note that all potentially hard-to-estimate parameters are contained in $B.$
The main idea of this paper is to leverage the knowledge and sparsity of $A$ and $B$ within the RLS framework, reducing the number of parameters to estimate and potentially speeding up identification.

Let $T_\rms\in\BR_{>0}$ be the sampling time. Assume the body-fixed frame aligns with the body's principal axes, rendering $J$ diagonal\footnote{The method is generalizable to nondiagonal entries.}. Discretize $A$ and $B$ through zero-order hold (ZOH) as
\begin{align}
    A_d & = e^{AT_\rms} = 
    \left[
    \begin{smallmatrix}
        1 & 0 & 0 & 0 & 0 & -\frac{T_\rms^2 g}{2} & T_\rms &  0 &  0 & 0 & -\frac{T_\rms^3 g}{6} &  0 \\
        0 & 1 & 0 & 0 & \frac{T_\rms^2 g}{2} &  0 &  0 & T_\rms &  0 & \frac{T_\rms^3 g}{6} &  0 &  0 \\
        0 & 0 & 1 & 0 & 0 &  0 &  0 &  0 & T_\rms & 0 &  0 &  0 \\
        0 & 0 & 0 & 1 & 0 &  0 &  0 &  0 &  0 & 0 &  0 & T_\rms \\
        0 & 0 & 0 & 0 & 1 &  0 &  0 &  0 &  0 &T_\rms &  0 &  0 \\
        0 & 0 & 0 & 0 & 0 &  1 &  0 &  0 &  0 & 0 & T_\rms &  0 \\
        0 & 0 & 0 & 0 & 0 &  -T_\rms g &  1 &  0 &  0 & 0 & -\frac{T_\rms^2 g}{2} &  0 \\
        0 & 0 & 0 & 0 &  T_\rms g &  0 &  0 &  1 &  0 & \frac{T_\rms^2 g}{2} &  0 &  0 \\
        0 & 0 & 0 & 0 & 0 &  0 &  0 &  0 &  1 & 0 &  0 &  0 \\
        0 & 0 & 0 & 0 & 0 &  0 &  0 &  0 &  0 & 1 &  0 &  0 \\ 
        0 & 0 & 0 & 0 & 0 &  0 &  0 &  0 &  0 & 0 &  1 &  0 \\
        0 & 0 & 0 & 0 & 0 &  0 &  0 &  0 &  0 & 0 &  0 &  1
    \end{smallmatrix}
    \right], \nonumber \\
    B_\rmd & =\int_0^{T_\rms} e^{A\tau}d\tau B = 
    \left[
    \begin{smallmatrix}
         0 & 0 & \ast & 0 \\
         0 & \ast & 0 & 0 \\
    \ast & 0 & 0 & 0 \\
         0 & 0 & 0 & \ast \\
         0 & \ast & 0 & 0 \\
         0 & 0 & \ast & 0 \\
         0 & 0 & \ast & 0 \\
         0 & \ast & 0 & 0 \\
    \ast & 0 & 0 & 0 \\
         0 & \ast & 0 & 0 \\
         0 & 0 & \ast & 0 \\
         0 & 0 & 0 & \ast \\
    \end{smallmatrix}
    \right]. \label{eq:discrete_sys}
\end{align}
Note that, since $A$ is non-invertible, the ZOH discretization of $B$ involves the computation of an integral, which must be numerically approximated.
Because the closed-form solution is difficult to obtain, we determine the canonical form of $B_\rmd$ with `$\ast$' as placeholders for nonzero elements.

Let $y_k = x_k + \nu_k$ be the measured state at step $k,$ where $\nu_k\in\BR^{12}$ is the sensor noise.
For all $k\ge 0,$ define the RLS performance variable as
\begin{align}
    z_k \triangleq y_k - \hat{y}_k, \label{eq:rls_perf1}
\end{align}
where $\hat{y}_k \triangleq A_\rmd y_{k-1} + \hat{B}_{\rmd,k} u_{k-1}$ is the state estimate at step $k,$ and $\hat{B}_{\rmd,k}$ is the estimate of $B_\rmd$ at step $k.$
Next, for all $k\ge 0,$ the regressor matrix $\Phi_k \in \BR^{12\times12}$ is given by
\begin{align}
    \Phi_k = \text{diag}\{u_3, u_2, u_1, u_4, u_2, u_3, u_3, u_2, u_1, u_2, u_3, u_4\}. \label{eq:regressor}
\end{align}
The diagonal element of the regressor is formed by traversing $B_\rmd$ row-wise and recording the column index of each nonzero element, which corresponds to the control index placed on the diagonal.

Using \eqref{eq:regressor} in \eqref{eq:rls_perf1} yields
\begin{align}
    z_k(\hat{\theta}) = y_k - A_\rmd y_{k-1} - \Phi_k\hat{\theta}, \label{eq:z_perf}
\end{align}
where $\hat{\theta}\in\BR^{12}$ is the coefficient vector to estimate, and the elements of $\hat{\theta}$ correspond to the placeholders `$\ast$' in \eqref{eq:discrete_sys}, arranged in row-major order.
%

Next, for all $k\ge 0,$ minimize the cumulative cost function
\begin{align}
    J_k(\hat{\theta}) = \sum_{i=0}^k \frac{\rho_i}{\rho_k}z_i^\rmT(\hat{\theta})z_i(\hat{\theta}) + \frac{1}{\rho_k}(\hat{\theta}-\theta_0)^\rmT P_0^{-1} (\hat{\theta}-\theta_0), \label{eq:RLS_cost}
\end{align}
where $\rho_k\triangleq \prod_{j=0}^k \lambda_j^{-1}\in\BR,$ $\lambda_k\in(0,1]$ is the forgetting factor, $P_0\in\RR{12}_{>0},$ and $\theta_0\in\BR^{12}$ is the initial estimate of the coefficient vector.
Note that, with \eqref{eq:z_perf}, the cost function \eqref{eq:RLS_cost} is convex and quadratic and, thus, has a unique global minimizer, which is updated by RLS as
\begin{align}
    \theta_{k+1} = \arg\min_{\hat{\theta}}J_k(\hat{\theta}). \label{eq:RLS}
\end{align}

\section{Predictive Cost Adaptive Control}\label{sec:pcac}

For real-time implementation, the control $u_k$ is computed between steps $k-1$ and $k,$ then applied at step $k$ and held constant until step $k+1.$

In particular, at each step $k>0,$ PCAC computes $u_{k+1}$ by reconstructing $\hat{B}_{\rmd,k}$ through RLS. 
The nonzero elements of the matrix, whose structure is defined by \eqref{eq:discrete_sys}, are populated by $\theta_{k+1},$ computed by \eqref{eq:RLS} and available at step $k.$

Next, using the one-step ahead output
$y_{1|k} \triangleq A_{\rmd} y_k + \hat{B}_{\rmd,k} u_k,$ PCAC propagates the identified discrete model over the horizon $\ell\ge1$ as
\begin{align}
    Y_{1|k,\ell} = \Gamma_{\ell} y_{1|k} + \hat{T}_{k,\ell} U_{1|k,\ell},
\end{align}
where
\begin{align}
    Y_{1|k,\ell} \triangleq \left[
    \begin{matrix}
        y_{1|k} \\
        \vdots \\
        y_{\ell|k}
    \end{matrix}
    \right] \in \BR^{12\ell},
    \;\;\; U_{1|k,\ell} \triangleq 
    \left[
    \begin{matrix}
        u_{1|k} \\
        \vdots \\
        u_{\ell|k}
    \end{matrix}
    \right] \in \BR^{4\ell},
\end{align}
where, for $i=1,\ldots,\ell,$ $y_{i|k}\in\BR^{12}$ is the $i$-step ahead output and $u_{i|k}\in\BR^{4}$ the $i$-step computed control.
The matrices $\Gamma_{\ell}$ and $\hat{T}_{k,\ell}$ are given by
%
\begin{align}
    \Gamma_{\ell} & \triangleq
    \left[
    \begin{matrix}
        I_{12} \\
        A_\rmd \\
        \vdots \\
        A_\rmd^{\ell-1}
    \end{matrix}
    \right], \nonumber \\
    \hat{T}_{k,\ell} & \triangleq
    \left[
    \begin{smallmatrix}
        0_{12\times4} & \ldots & \ldots & \ldots & \ldots & \ldots & 0_{12\times4} \\
        \hat{H}_{k,1} & 0_{12\times4} & \ldots & \ldots & \ldots & \ldots & 0_{12\times4} \\
        \hat{H}_{k,2} & \hat{H}_{k,1} & 0_{12\times4} & \ldots & \ldots & \ldots & 0_{12\times4} \\
        \hat{H}_{k,3} & \hat{H}_{k,2} & \hat{H}_{k,1} & 0_{12\times4} & \ldots & \ldots & 0_{12\times4} \\
        \hat{H}_{k,4} & \hat{H}_{k,3} & \hat{H}_{k,2} & \ddots & \ddots & & \vdots \\
        \vdots & \vdots & \vdots & \ddots & \ddots & \ddots & 0_{12\times4} \\
        \hat{H}_{k,\ell-1} & \hat{H}_{k,\ell-2} & \hat{H}_{k,\ell-3} & \ldots & \hat{H}_{k,2} & \hat{H}_{k,1} & 0_{12\times4}
    \end{smallmatrix}
    \right],
\end{align}
where, for $i=1,\ldots,\ell-1,$ $\hat{H}_{k,i} \triangleq A_\rmd^{i-1}\hat{B}_{\rmd,k}\;\;\in\BR^{12\times4}.$
Note that, since $A_\rmd$ is fixed, $\Gamma_{\ell}$ can be computed ahead for speed.

\begin{figure*}[!ht]
    \centering
    \includegraphics[trim=0 10 50 0, clip, width=0.9\textwidth]{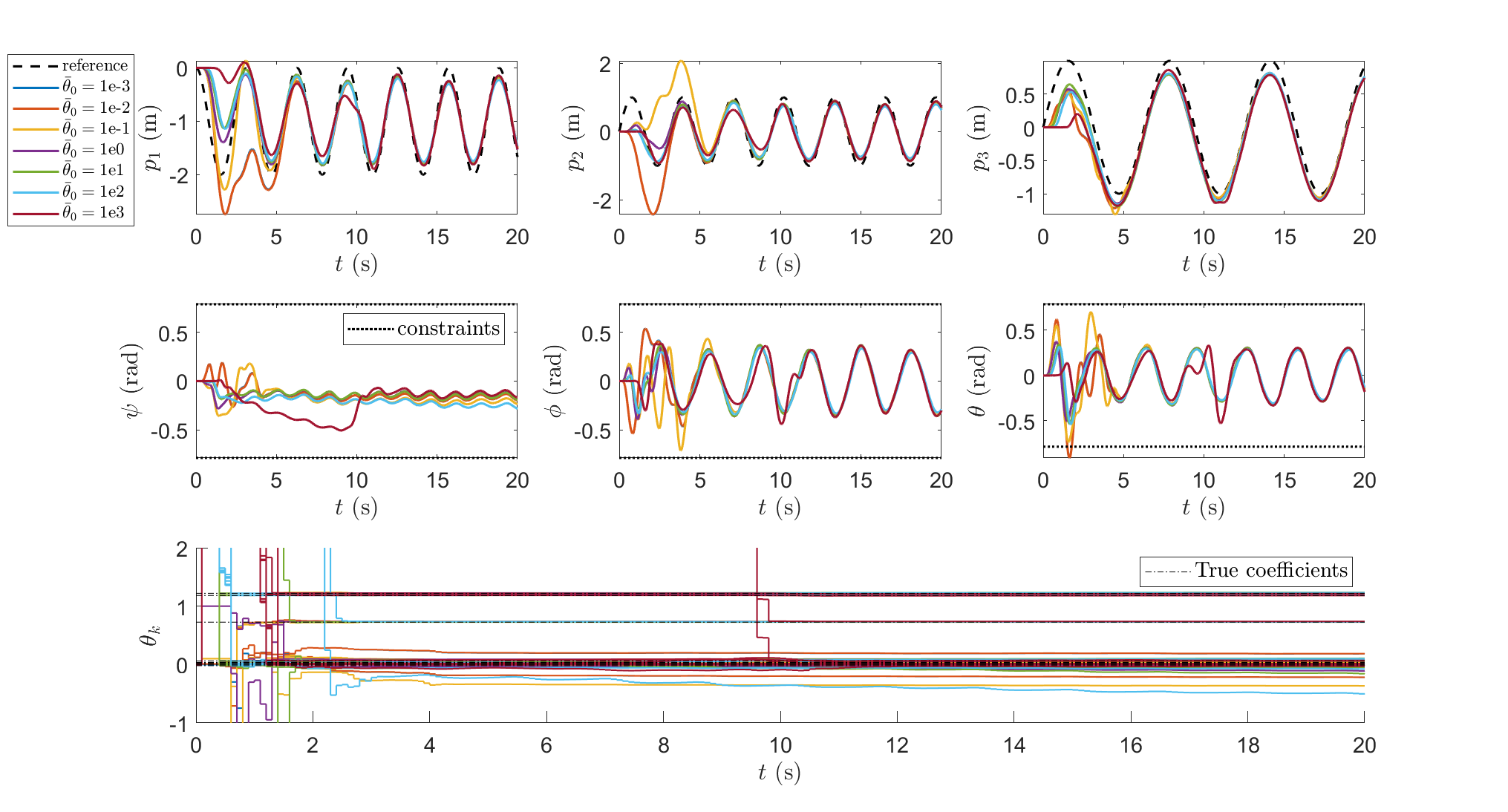}
    \caption{
    \textbf{Example 1}: Periodic trajectory tracking for various $\theta_0=[\bar{\theta}_0 \; \ldots \; \bar{\theta}_0]^\rmT.$ 
    In all cases, we can see from the top three subplots that, starting around $t=12$ s, the desired position trajectory $\vec{p}_\rmd$ is closely tracked by $\vec{p}.$
    The periodic command is not perfectly tracked because the PCAC-identified model lacks an internal model of the command signal. 
    Note in the top-left subplot that the undershoot effect at $t=2$ s decreases as $\bar{\theta}_0$ increases.
    Additionally, note that the Euler angles and the identified parameters contained in $\theta_k$ experience the longest transient duration for $\bar{\theta}_0=1\text{e3},$ which is approximately 10 s.
    For $\bar{\theta}=1\text{e-3}$ and $\bar{\theta}=1\text{e-2},$ the roll angle $\theta$ shortly violates the constraint at approximately $t=1.5$ s, causing the slack variable to become active.
    Constraint violation occurs at approximately $t = 2$ s because the identified model is still in transient, hindering PCAC's ability to enforce constraints.
    Moreover, note that, for all cases, although the parameters contained in $\theta_k$ do not converge exactly to the true values, they come sufficiently close to enable PCAC to use an adequately accurate model for control.
    }
    \label{fig:example1_state}
\end{figure*}

Next, let $r_k\in\BR^{12}$ be the state command at step $k,$ and define $\Delta u_{1|k} \triangleq u_{1|k}-u_k$ and, for $i > 1,$ $\Delta u_{i|k} \triangleq u_{i|k}-u_{i-1|k}.$
PCAC solves the QP-based receding-horizon optimization
\begin{align}
    \min_{u_{1|k},\ldots,u_{\ell|k},\varepsilon_1,\ldots,\varepsilon_\ell} & \sum_{i=1}^\ell (y_{i|k} - r_i)^\rmT \SQ_i (y_{i|k} - r_i)  + \nonumber \\
    & \quad \quad (\Delta u_{i|k})^\rmT \SR \Delta u_{i|k} + \varepsilon_i^\rmT \mathcal{S} \varepsilon_i, \nonumber \\
    \text{subject to:} \quad \quad 
 & \SC y_{i|k} + \SD \le \varepsilon_i, \quad i = 1,\ldots,\ell, \nonumber \\
 & u_{\text{min}} \le u_{i|k} \le u_{\text{max}}, \quad i = 1,\ldots,\ell, \nonumber \\
 & \Delta u_{\text{min}} \le \Delta u_{i|k} \le \Delta u_{\text{max}}, \quad i = 1,\ldots,\ell, \nonumber \\
 & 0_{n_\rmc\times1} \le \varepsilon_i, \quad i = 1,\ldots,\ell, \label{eq:QP}
\end{align}
where the inequalities apply element-wise, $\varepsilon_i\in\BR^{n_\rmc}$ is the slack variable at step $i,$ $n_\rmc \ge 0$ the number of constraints, $\SQ_i\in\RR{12}_{>0}$ the cost-to-go output weight at step $i,$ $\SR\in\RR{4}_{\ge0}$ the control weight, $\mathcal{S}\in\RR{n_\rmc}_{\ge0}$ the constraint relaxation weight, $\SC\in\BR^{n_\rmc\times12},$ $\SD\in\BR^{n_\rmc},$ $u_{\text{min}}\in\BR^{4},$ $u_{\text{max}}\in\BR^{4},$ $\Delta u_{\text{min}}\in\BR^{4},$ and $\Delta u_{\text{max}}\in\BR^{4}.$
Slacks are introduced to ensure the problem remains feasible at all times.

Next, to compute $u_{k+1},$ add a gravity compensation term to $u_{1|k}$ as
\begin{align}
    u_{k+1} = u_{1|k} + [\begin{matrix} mg & 0 & 0 & 0 \end{matrix}]^\rmT.
\end{align}
This ensures that PCAC operates near hover points and uses a valid linearized model to facilitate identification.

%

\section{Parameter Estimation Accuracy and Persistence of Excitation}

In this work, parameter estimation is driven by the closed-loop interaction between control and identification, without the use of designed excitation signals. The system excites itself through the control process, allowing parameters to be refined online via the intrinsic duality between control and identification \cite{feldbaum1963dual}. While persistence of excitation (P.E.) is not enforced explicitly, it is observed empirically that the evolving control inputs induce sufficient excitation over time to enable convergence of key parameters. In practice, the identified model achieves tracking and constraint satisfaction within acceptable limits, indicating that parameter accuracy is sufficient for control performance in the presented simulations.

\section{Simulations}\label{sec:simulations}

At each step $k\ge0,$ \eqref{eq:nonlinear_system} is simulated using Matlab's ODE45 over the interval $t\in[kT_\rms,(k+1)T_\rms].$
The final state at $t=(k+1)T_\rms$ serves as the initial condition for the next simulation.

PCAC implements a real-time RLS, as described by \cite{8716768}, employing a variable-rate forgetting (VRF) factor, as proposed by \cite{BRUCE2020109052}, to facilitate identification in the case of abrupt changes.
The solution of \eqref{eq:QP} is computed using a warm start.

Let $m=4.34$ kg, $J=\text{diag}\{0.082,0.0845,0.1377\}$ kg$.$m$^2,$ and $T_\rms = 0.1$ s.
The initial conditions are $\vec{p}(0)=[0 \; 0 \; 0]^\rmT,$ $\dot{\vec{p}}(0) = [0 \; 0 \; 0]^\rmT,$ $\vec{\xi}(0) = [0 \; 0 \; 0]^\rmT,$ and $\vec{\omega}(0) = [0 \; 0 \; 0]^\rmT,$ and the input constraints are $u_{\text{max}} = -u_{\text{min}}$ and $\Delta u_{\text{max}} = - \Delta u_{\text{min}},$ where $u_{\text{max}} = [20 \; 2 \; 2 \; 2]^\rmT$ and $\Delta u_{\text{max}} = [5 \; 0.3 \; 0.3 \; 0.3]^\rmT.$
The Euler angles are constrained by $-\vec{\xi}_{\text{max}} \le \vec{\xi} \le \vec{\xi}_{\text{max}},$ where $\vec{\xi}_{\text{max}} = [\frac{\pi}{4} \; \frac{\pi}{4} \; \frac{\pi}{4}]^\rmT,$ ensuring that the quadrotor operates near hover points.

\begin{figure*}[!ht]
    \centering
    \includegraphics[trim=0 20 40 0, clip, width=0.9\linewidth]{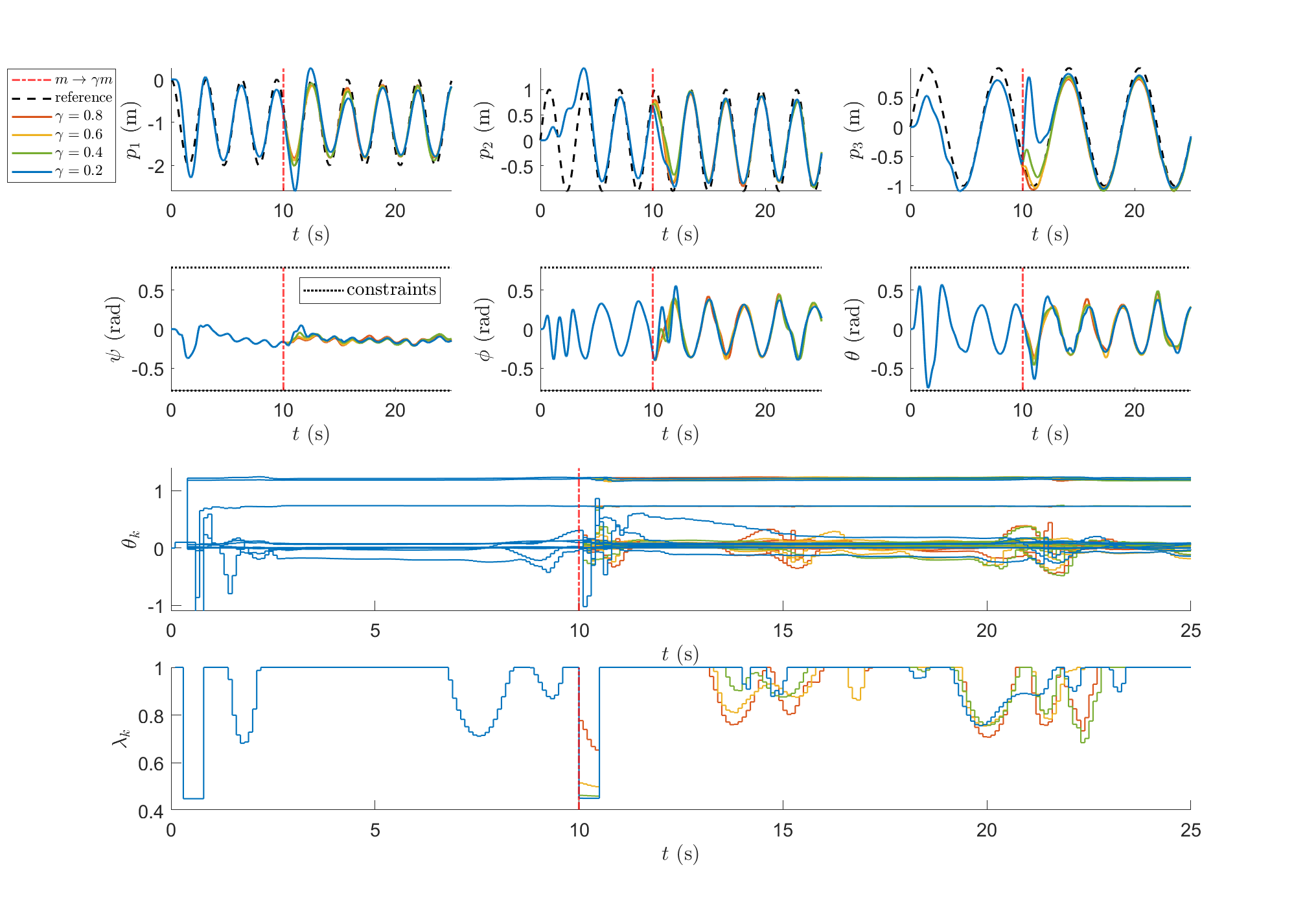}
    \caption{\textbf{Example 2}: Periodic trajectory tracking with an abrupt change in quadrotor mass at $t=10$ s.
    Note from the top three subplots that, in all cases, after a transient period of 5 s starting at $t=10$ s, the periodic trajectory is closely tracked by $\vec{p}.$
    Furthermore, note that, at $t=10$ s, $\lambda_k$ drops to approximately 0.45, triggering forgetting and facilitating re-identification.
    }
    \label{fig:example2}
\end{figure*}

PCAC utilizes VRF with $P_0 = 10^6I_{12},$ $\tau_{\mathrm{n}} = 5,$ and $\tau_\rmd = 25.$
Furthermore, let $\ell=10,$ $\SQ_i = \text{diag}\{50,50,50,10,10,10,50,50,50,10,10,10\}$ for all $i < \ell,$ $\SQ_\ell = 10 Q_1,$ $\SR = \text{diag}\{0.1,0.1,0.1,\allowbreak0.1\},$ and $\mathcal{S} = 10^6 I_3.$

The periodic position command is set to $p_{\rmd}(t) = [\cos(2t)-1 \allowbreak \; \sin(2t) \; \sin(t)]^\rmT,$ generating the state command
$r(t) = [\cos(2t)-1 \; \sin(2t) \; \sin(t) \; 0 \; 0 \; 0 \; \allowbreak -2\sin(2t) \; 2\cos(2t) \; \cos(t) \; 0 \; 0 \; 0]^\rmT.$
This command trajectory is chosen to be continuously differentiable and periodic to facilitate steady-state tracking analysis under dynamic commands.

\subsection{Example 1: Periodic Trajectory Tracking for Various $\theta_0$}

Let $\eta = 1\text{e-3}$ and $\vec{p}=[p_1 \; p_2 \; p_3]^\rmT.$ Figure \ref{fig:example1_state} shows the results for various $\theta_0=[\bar{\theta}_0 \; \ldots \; \bar{\theta}_0]^\rmT\in\BR^{12},$ demonstrating the speed with which RLS can identify the system from different initial conditions.

Specifically, note that, although excitation is required at startup for identification, the known model structure reduces the number of parameters to be identified. As a result, RLS can identify the system with sufficient speed and accuracy, allowing PCAC to control the quadrotor.

\subsection{Example 2: Resilience to Abrupt Changes in Quadrotor Mass}

Let $\eta=0.99.$
For various values of $\gamma\in(0,1),$ consider an abrupt change in the quadrotor mass as
\begin{align}
m(t) = 
\begin{cases}
    4.34, & \text{for $t < 10$ s}, \\
    \gamma 4.34, & \text{for $t\ge 10$ s}.
\end{cases}
\end{align}

Figure \ref{fig:example2} shows the results for $\gamma=0.8,$ $\gamma=0.6,$ $\gamma=0.4,$ and $\gamma=0.2.$
VRF plays a crucial role in forgetting past data when there is an abrupt change in system parameters.
In essence, VRF compares two data windows.
If a mismatch is detected between the data and the estimates, VRF activates, promptly forgetting previous data to readapt the parameters used in PCAC.

\section{Conclusion}\label{sec:conclusion}

This paper has presented a centralized predictive cost adaptive control (PCAC) strategy for the position and attitude control of quadrotors.
By transforming the $SO(3)$ kinematics into Euler's, we derived a linearized system around hover points.
Next, extracting the $A$ and $B$ matrices, we observed that the elements within $A$ are known a priori, while uncertainties reside in $B.$
Based on this, we proposed a recursive-least-squares (RLS) approach that leverages the linearized system structure and unique sparsity, enabling the reconstruction of $\hat{B}_{\rmd,k}$ at each step $k$ for prediction.

To demonstrate the rapid identification capabilities of RLS, we provided two examples.
The first illustrated the robustness of PCAC with varying initial conditions, and the second showed its resilience to abrupt changes in quadrotor mass.

Future works will focus on enhancing PCAC by incorporating exogenous disturbance estimation, improving adaptability to mass changes in the gravity compensation term.

\section*{Acknowledgment}

The author thanks Prof. Bernstein for initiating this research direction.

\bibliographystyle{IEEETran.bst}
\bibliography{bibliography.bib}

\begin{thebibliography}{10}
\providecommand{\url}[1]{#1}
\csname url@rmstyle\endcsname
\providecommand{\newblock}{\relax}
\providecommand{\bibinfo}[2]{#2}
\providecommand\BIBentrySTDinterwordspacing{\spaceskip=0pt\relax}
\providecommand\BIBentryALTinterwordstretchfactor{4}
\providecommand\BIBentryALTinterwordspacing{\spaceskip=\fontdimen2\font plus
\BIBentryALTinterwordstretchfactor\fontdimen3\font minus
  \fontdimen4\font\relax}
\providecommand\BIBforeignlanguage[2]{{%
\expandafter\ifx\csname l@#1\endcsname\relax
\typeout{** WARNING: IEEEtran.bst: No hyphenation pattern has been}%
\typeout{** loaded for the language `#1'. Using the pattern for}%
\typeout{** the default language instead.}%
\else
\language=\csname l@#1\endcsname
\fi
#2}}

\bibitem{rs11121443}
\BIBentryALTinterwordspacing
H.~Yao, R.~Qin, and X.~Chen, ``{U}nmanned {A}erial {V}ehicle for remote sensing
  applications—a review,'' \emph{Remote Sensing}, vol.~11, no.~12, 2019.
  [Online]. Available: \url{https://www.mdpi.com/2072-4292/11/12/1443}
\BIBentrySTDinterwordspacing

\bibitem{8827409}
J.~Kim, S.~A. Gadsden, and S.~A. Wilkerson, ``A comprehensive survey of control
  strategies for autonomous quadrotors,'' \emph{Canadian Journal of Electrical
  and Computer Engineering}, vol.~43, no.~1, pp. 3--16, 2020.

\bibitem{idrissi2022review}
M.~Idrissi, M.~Salami, and F.~Annaz, ``A review of quadrotor unmanned aerial
  vehicles: applications, architectural design and control algorithms,''
  \emph{Journal of Intelligent \& Robotic Systems}, vol. 104, no.~2, p.~22,
  2022.

\bibitem{6196930}
S.~Gupte, P.~I.~T. Mohandas, and J.~M. Conrad, ``A survey of quadrotor unmanned
  aerial vehicles,'' in \emph{2012 Proceedings of IEEE Southeastcon}, 2012, pp.
  1--6.

\bibitem{aerospace8120363}
\BIBentryALTinterwordspacing
N.~Elmeseiry, N.~Alshaer, and T.~Ismail, ``A detailed survey and future
  directions of unmanned aerial vehicles ({UAV}s) with potential
  applications,'' \emph{Aerospace}, vol.~8, no.~12, 2021. [Online]. Available:
  \url{https://www.mdpi.com/2226-4310/8/12/363}
\BIBentrySTDinterwordspacing

\bibitem{szafranski2011different}
G.~Szafranski and R.~Czyba, ``Different approaches of pid control {UAV} type
  quadrotor,'' in \emph{International Micro Air Vehicle conference and
  competitions 2011}.\hskip 1em plus 0.5em minus 0.4em\relax 't Harde, The
  Netherlands: Delft University of Technology and Thales, 2011.

\bibitem{PAREDES2021372}
\BIBentryALTinterwordspacing
J.~Paredes, P.~Sharma, B.~Ha, M.~Lanchares, E.~Atkins, P.~Gaskell, and
  I.~Kolmanovsky, ``Development, implementation, and experimental outdoor
  evaluation of quadcopter controllers for computationally limited embedded
  systems,'' \emph{Annual Reviews in Control}, vol.~52, pp. 372--389, 2021.
  [Online]. Available:
  \url{https://www.sciencedirect.com/science/article/pii/S1367578821000420}
\BIBentrySTDinterwordspacing

\bibitem{9144371}
P.~Zheng, X.~Tan, B.~B. Kocer, E.~Yang, and M.~Kovac, ``Tiltdrone: A
  fully-actuated tilting quadrotor platform,'' \emph{IEEE Robotics and
  Automation Letters}, vol.~5, no.~4, pp. 6845--6852, 2020.

\bibitem{NGUYEN2021109586}
\BIBentryALTinterwordspacing
T.~W. Nguyen, M.~Hosseinzadeh, and E.~Garone, ``Thrust vector control of
  constrained multibody systems,'' \emph{Automatica}, vol. 129, p. 109586,
  2021. [Online]. Available:
  \url{https://www.sciencedirect.com/science/article/pii/S0005109821001060}
\BIBentrySTDinterwordspacing

\bibitem{WANG2023108088}
\BIBentryALTinterwordspacing
J.~Wang, K.~A. Alattas, Y.~Bouteraa, O.~Mofid, and S.~Mobayen, ``Adaptive
  finite-time backstepping control tracker for quadrotor {UAV} with model
  uncertainty and external disturbance,'' \emph{Aerospace Science and
  Technology}, vol. 133, p. 108088, 2023. [Online]. Available:
  \url{https://www.sciencedirect.com/science/article/pii/S1270963822007623}
\BIBentrySTDinterwordspacing

\bibitem{9612636}
T.~W. Nguyen, S.~A.~U. Islam, D.~S. Bernstein, and I.~V. Kolmanovsky,
  ``{P}redictive {C}ost {A}daptive {C}ontrol: A numerical investigation of
  persistency, consistency, and exigency,'' \emph{IEEE Control Systems
  Magazine}, vol.~41, no.~6, pp. 64--96, 2021.

\bibitem{9482724}
T.~W. Nguyen, I.~V. Kolmanovsky, and D.~S. Bernstein, ``Sampled-data
  output-feedback model predictive control of nonlinear plants using online
  linear system identification,'' in \emph{2021 American Control Conference
  (ACC)}, 2021, pp. 4692--4697.

\bibitem{doi:10.2514/6.2025-0033}
\BIBentryALTinterwordspacing
J.~C.~V. Schaaf, Q.~Lu, H.~Kumar, B.~Ozmadenci, K.~Fidkowski, and D.~Bernstein,
  ``{P}redictive {C}ost {A}daptive {C}ontrol of in-ground-effect flight using
  active flow control with unmodeled, unsteady aerodynamics,'' in \emph{AIAA
  SCITECH 2025 Forum}, 2025. [Online]. Available:
  \url{https://arc.aiaa.org/doi/abs/10.2514/6.2025-0033}
\BIBentrySTDinterwordspacing

\bibitem{doi:10.2514/6.2025-2081}
\BIBentryALTinterwordspacing
R.~Richards, J.~Paredes, and D.~Bernstein, ``{P}redictive {C}ost {A}daptive
  {C}ontrol of fixed-wing aircraft without prior modeling,'' in \emph{AIAA
  SCITECH 2025 Forum}, 2025. [Online]. Available:
  \url{https://arc.aiaa.org/doi/abs/10.2514/6.2025-2081}
\BIBentrySTDinterwordspacing

\bibitem{doi:10.2514/1.G008859}
\BIBentryALTinterwordspacing
R.~J. Richards, S.~A.~U. Islam, and D.~S. Bernstein, ``{P}redictive {C}ost
  {A}daptive {C}ontrol of the {NASA} benchmark flutter model,'' \emph{Journal
  of Guidance, Control, and Dynamics}, pp. 1--17, 2025. [Online]. Available:
  \url{https://doi.org/10.2514/1.G008859}
\BIBentrySTDinterwordspacing

\bibitem{feldbaum1963dual}
A.~A. Feldbaum, ``Dual control theory problems,'' \emph{IFAC Proceedings
  Volumes}, vol.~1, no.~2, pp. 541--550, 1963.

\bibitem{8716768}
S.~A.~U. Islam and D.~S. Bernstein, ``Recursive least squares for real-time
  implementation [lecture notes],'' \emph{IEEE Control Systems Magazine},
  vol.~39, no.~3, pp. 82--85, 2019.

\bibitem{BRUCE2020109052}
\BIBentryALTinterwordspacing
A.~L. Bruce, A.~Goel, and D.~S. Bernstein, ``Convergence and consistency of
  recursive least squares with variable-rate forgetting,'' \emph{Automatica},
  vol. 119, p. 109052, 2020. [Online]. Available:
  \url{https://www.sciencedirect.com/science/article/pii/S0005109820302508}
\BIBentrySTDinterwordspacing

\end{thebibliography}

\end{document}